# Chip-based photonic radar for high-resolution imaging

SIMIN LI, ZHENGZE CUI, XINGWEI YE, JING FENG, YUE YANG, ZHENGQIANG HE, RONG CONG, DAN ZHU, FANGZHENG ZHANG, SHILONG PAN*

*Key Laboratory of Radar Imaging and Microwave Photonics, Ministry of Education, Nanjing University of Aeronautics and Astronautics, Nanjing 210016, China*
*Corresponding author: pans@nuaa.edu.cn*

**Radar is the only sensor that can realize target imaging at all time and all weather, which would be a key technical enabler for future intelligent society. Poor resolution and large size are two critical issues for radar to gain ground in civil applications. Conventional electronic radars are difficult to address both issues especially in the relatively low-frequency band. In this work, we propose and experimentally demonstrate, for the first time to the best of our knowledge, a chip-based photonic radar based on silicon photonic platform, which can implement high resolution imaging with very small footprint. Both the wideband signal generator and the de-chirp receiver are integrated on the chip. A broadband photonic imaging radar occupying the full Ku band is experimentally established. A high precision range measurement with a resolution of 2.7 cm and an error of less than 2.75 mm is obtained. Inverse synthetic aperture (ISAR) imaging of multiple targets with complex profiles are also implemented.**

With the unique capability of target imaging at all time and all weather, imaging radar will play an increasingly important role for future intelligent society [1, 2]. Chip-based imaging radars are critical for miniaturized platforms such as unmanned aerial vehicles (UAVs), autonomous vehicles or even mobile devices to navigate and sense the environment, due to the great benefits on the reductions of cost, power consumption, volume, and weight. With the fast development of the monolithic microwave integrated circuit (MMIC) technology, a number of chip-based radars have been demonstrated [3-7]. For radar imaging, a waveform with a large bandwidth is usually required to achieve a high resolution because the range resolution is inversely proportional to the bandwidth. However, it is extremely challenging for electronic chip-based radars to generate and process broadband waveforms especially in the low-frequency band in which high power volume, high power efficiency, and long-distance detection can be easily achieved. Ref. [7] reported a Ku-band on-chip radar transmitter with a recorded bandwidth of 1.48 GHz for UAV application. The resolution is about 11.2 cm, which is insufficient for the detection of small targets.

On the other hand, photonic signal generation and processing provides distinct features in terms of wide frequency coverage, broad instantaneous bandwidth, low-frequency dependent loss, and immunity to electromagnetic interference, which could be a key technical enabler for future ultrahigh resolution and real-time imaging radar. In 2014, P. Ghelfi et al. proposed the world-first photonics-based coherent radar by employing a mode-locked laser (MLL) to realize both signal generation and sampling [8]. After that, different photonic radar architectures have been demonstrated. In the transmitter, large bandwidth linear frequency modulated (LFM) signals were generated by optical frequency multiplication [9-11], optical heterodyning [12] or photonic digital-to-analog converter [13]. In the receiver, the echoes are handled by photonic time-stretched processing [9, 12], photonic de-chirping [10, 13], or photonic balanced I/Q de-chirping [14]. Ultrahigh resolution radar imaging up to 1.3 cm was reported [15]. Application scenarios of the photonic radars have also been investigated, such as aircraft/UAV/maritime traffic detection [8, 9, 16, 17], automotive sensing [18], and landslides monitoring [19].

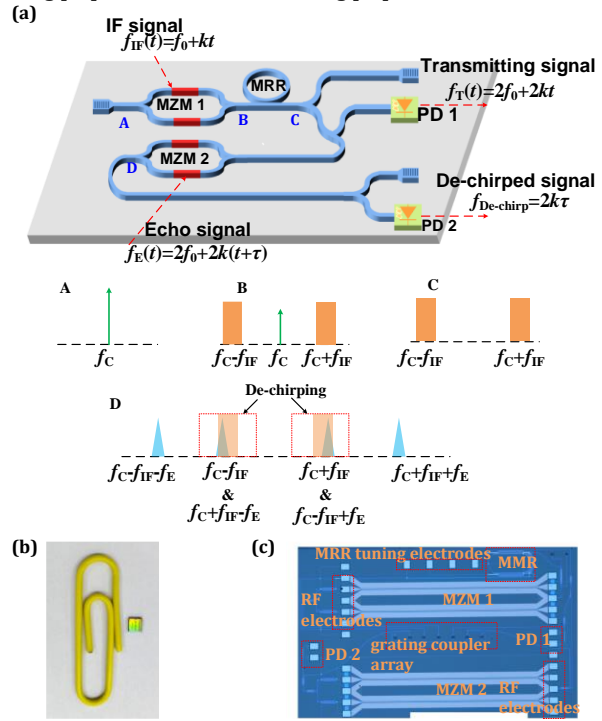

Fig. 1. (a) Schematic diagram of the proposed chip-based photonic radar. MZM: Mach-Zehnder modulator; MRR: micro-ring resonator; PD: photodetector. (b) The picture and (c) zoom-in view of the fabricated chip.

Although the existed achievements have demonstrated the significantly improved performance and showed the great potential of the photonic radar, all of the reported photonic radars are constructed based on discrete optical components, leading to a bulky system with low reliability. Thanks to the long-standing efforts devoted to the photonic integration technologies, chip-based photonic radar with a dramatic reduction in the footprint is becoming possible [20]. Many integrated microwave photonic subsystems were reported previously, including the integrated microwave photonic filter [21], microwave photonic true time delay line [22], optical beamformer [23], integrated optoelectronic oscillator [24], programmable photonic signal processor chip [25] and so on.

In this paper, we demonstrate, for the first time to the best of our knowledge, a chip-based photonic imaging radar based on silicon photonic platform. Both the photonic frequency-doubled LFM signal generator and the photonic de-chirp receiver are integrated on a chip with a footprint of 1.45 mm×2.5 mm. In a proof-of-concept experiment, a photonic radar with a bandwidth covering the full Ku-band (12-18 GHz) is established. A high precision distance measurement with an error of less than 2.75 mm is implemented, and inverse synthetic aperture (ISAR) imaging with a resolution of 2.7 cm is achieved. The resolution is 4 times better than the on-chip electrical radar in Ref [7].

Figure 1 shows the schematic of the integrated photonic radar chip. An optical carrier with a frequency of $f_c$ from a tunable laser source is coupled into the chip via a grating coupler, which is then sent into an on-chip Mach-Zehnder modulator (MZM1). MZM1 is driven by an intermediate-frequency (IF) LFM signal with an expression of $f_{IF}(t)=f_0+kt$, and is biased at the lowest transmission point to suppress the optical carrier. A micro-ring resonator (MRR) filter is followed to enhance the carrier suppression. Ideally, only the ±1st-order sidebands with frequencies of $f_{MZM1}(t)=f_c±(f_0+kt)$ would be output from the MRR. Then the optical signal is separated into two paths by a 3-dB optical coupler. In the upper path, the optical signal is sent to a photodetector (PD1) to generate the frequency-doubled LFM signal, which is expressed by $f_T(t)=2f_0+2kt$. The frequency-doubled LFM signal is amplified by an electrical amplifier and emitted through an antenna to the free space. In the lower path, the optical signal is employed as the local signal for de-chirp processing of the received signal. The received signal is amplified by a low noise amplifier (LNA) and sent to the RF port of a second MZM (MZM2). For the simplest scenario, the echo signal is a replica of the transmitting signal with a time delay of $\tau$ which can be written as $f_E(t)= 2f_0+2k(t+\tau)$. Then, the optical signal output from MZM2, which is expressed as $f_{MZM2}(t)=f_c±(f_0+kt) ±(2f_0+2kt+2k\tau)$, is sent to another PD (PD2) to obtain the de-chirped signal with a frequency of $f_{de\text{-}chirp}(t)= 2k\tau$. As a result, the distance of the target can be calculated by [10]

$$L = \frac{cT}{2B} f_{de\text{-}chirp} \quad (1)$$

where $c$ is the velocity of light, $T$ and $B$ are the bandwidth and pulse width of the generated LFM signal. Due to the fact that low frequency de-chirped signal can be processed by a low-speed ADC, real-time ISAR imaging for moving targets can be easily achieved. The range resolution of the radar is determined by the 3-dB bandwidth of the de-chirped signal $\Delta f_{3dB}$, i.e.,

$$L_{RES} = \frac{cT}{2B} \Delta f_{3dB} \quad (2)$$

For an ideal case, $\Delta f_{3dB}=1/T$, so the best resolution that can be achieved is $L_{RES}=c/2B$.

The proposed chip is fabricated in imec's silicon photonics platform (iSIPP50G). Figure 1 (b) and (c) show the images of the chip captured by a camera and a microscope, respectively. The footprint of the chip-based photonic radar is about 1.45 mm×2.5 mm. The MRR is designed as the racetrack shape with a perimeter of 996.6 μm. The MZMs and PDs are extracted from the device library of iSIPP50G. In order to monitor the optical spectral response, some optical output ports are introduced. The light is coupled into and out of the chip using a fiber array through a grating coupler array with an interval of 127 μm. The measured fiber to fiber insertion loss is about 10 dB.

To investigate the performance of the chip-based photonic radar, an experiment is carried out. A continuous-wave (CW) light with a wavelength of 1552.9 nm is coupled into the chip via a polarization controller. An arbitrary waveform generator (Tektronix AWG70000) produces an IF-LFM signal with a bandwidth of 3 GHz (6-9 GHz) and a pulse width of 100 μs. The IF-LFM signal and a direct current (DC) bias signal are combined by a bias-tee and then fed into a microwave probe which is connected to the RF electrode of MZM1. By adjusting the DC bias and tuning the MRR's resonator wavelength to align the optical carrier, a double-sideband modulated optical signal with the carrier suppressed is obtained, which is then coupled out of the chip and sent into a PD to perform the optical-to-electrical conversion. Amplified by a low-noise amplifier (LNA) and filtered by an electrical band-pass filter, the frequency-doubled LFM signal is emitted to the free space by a horn antenna. The echo signal collected by a second horn antenna is amplified by another LNA and then launched to drive MZM2 on the chip. The output optical signal is detected by an external low-speed PD, and the generated electrical signal is sampled by a real-time oscilloscope (Keysight 93304).

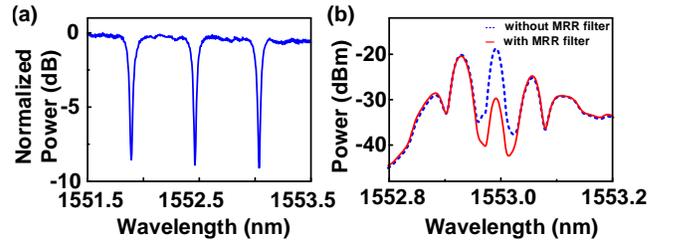

Fig. 2. (a) Transmission response of the MRR. (b) The spectra of the optical DSB-SC signal without (blue dash line) and with (red solid line) the incorporation of the MRR filter.

Figure 2 (a) shows the measured transmission response of the MRR, which has a free-spectral range (FSR) of 0.57 nm and an extinction ratio of 9 dB. Figure 2 (b) illustrates the spectra of the double sideband suppressed-carrier (DSB-SC) modulation signal without and with the incorporation of the MRR filter. The best carrier suppression state for the on-chip MZM by adjusting the DC bias still exhibits a strong optical carrier, shown as the blue dashed line in Fig. 2 (b). A possible reason is that a phase difference between the two arms of the MZM is introduced by the free-carrier dispersion and the thermo effects, leading to an unideal DSB-SC modulation. When the resonator wavelength of the MRR is tuned to filter the optical carrier, the optical carrier is more than 5-dB lower than the sideband according to the red solid line in Fig. 2(b).

The generated frequency-doubled LFM transmitting signal is filtered and captured by the real-time oscilloscope with a sampling rate of 80-GSa/s. The spectrum and time-frequency relationship are calculated using fast Fourier transform (FFT) and short-time Fourier transform, respectively. The results are shown in Fig. 3. As can be seen, the transmitting signal covers the full Ku-band (12 to 18 GHz) with a bandwidth of 6 GHz. The pulse width is 100 μs. The magnitude variation of the signal is within ±1 dB and the out-of-band spurious rejection ratio is greater than 25 dB. It should be noted that frequency multiplication with a higher multiplication factor is also possible if the modulation index is increased, the DC bias is properly set and the wavelength of the MRR is carefully adjusted [26].

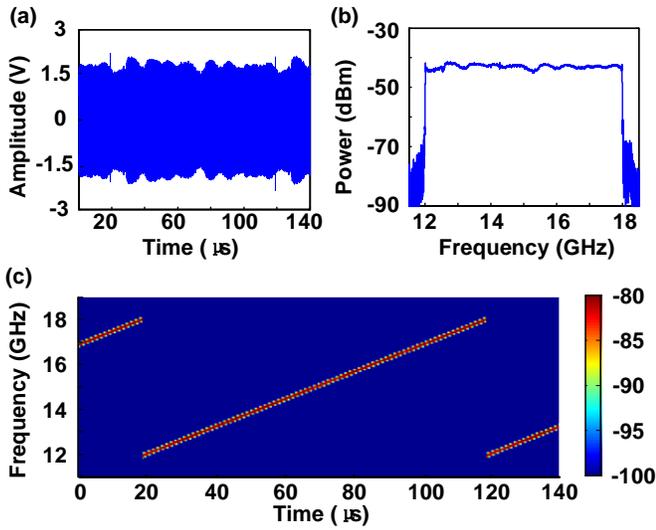

Fig. 3. The measured results of the generated frequency-doubled LFM signal. (a) The time-domain waveform, (b) the electrical spectrum and (c) the instantaneous frequency.

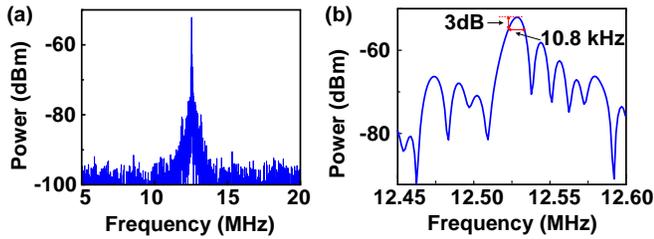

Fig. 4. The measured spectra of the de-chirped signal. (a) The spectrum in a frequency range of 15 MHz, (b) the zoom-in view of the spectrum around the main peak.

To evaluate the wideband de-chirp signal processing capability of the chip, the generated frequency-doubled LFM signal is modulated on MZM2 directly without emitting to the free space by the transmit antenna. The de-chirped signal is sampled by the real-time oscilloscope with a sampling rate of 100 MSa/s. Figure 4 shows the spectra performed by the FFT of the captured data with a time window of 100 μs. A low-frequency component at 12.5 MHz with a 3-dB bandwidth of 10.8 kHz is achieved. According to (2), the range resolution is about 2.7 cm, which is very close to the theoretical value of 2.5 cm determined by the signal bandwidth [10]. It should be noted that there is a sidelobe with a sidelobe suppression ratio (SLSR) of ~6 dB, 15.5-kHz apart from the main peak. This sidelobe should be caused by the RF crosstalk between the probes connected to the two MZMs on the chip, which could be suppressed by carefully designing the RF packaging.

A measurement of the target with a varying distance to the antennas is implemented to evaluate the range measurement accuracy. The target is placed about 30 cm away from the antennas initially and then the distance is increased to 46 cm with a step of 2 cm. Figure 5 shows the measured results. The measured frequency errors are kept within ±1.1 kHz, corresponding to a range error of less than 2.75 mm.

Then, the ISAR imaging experiment is implemented. The targets to be detected are placed on a turntable with a speed of about 360°/s. The de-chirped signal is sampled by the real-time oscilloscope with a sampling rate of 50 MSa/s. The ISAR images are constructed based on two-dimensional Fourier transformation. Firstly, two targets are placed on the turntable with range and cross-range distances of 3 cm and 10 cm, respectively. Fig. 6 (a) shows the obtained ISAR image. The two targets can be clearly distinguished, with the range and cross-range distances of 3.05 and 10.45 cm which agree well with the actual value. The image of multiple targets like a capital "A" constitute by 6 trihedral corner reflectors is shown in Fig. 6 (b), which also confirms the high resolution of the chip-based photonic radar. Finally, the ISAR image of an airplane model with a 28-cm body and a 32-cm wingspan is also obtained, as shown in Fig. 6(c).

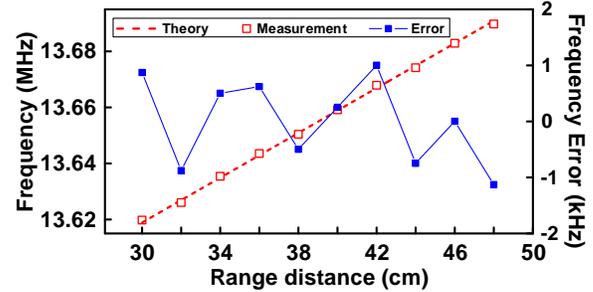

Fig. 5. The measured frequency of the de-chirped signal for a target moved from 30 to 46 cm with a step of 2 cm.

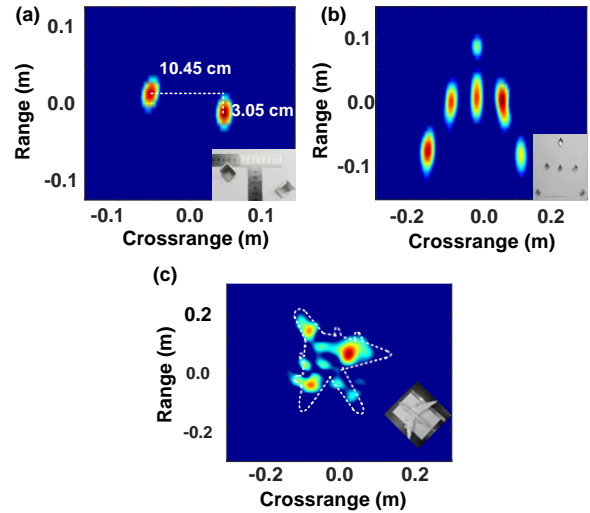

Fig. 6. ISAR images of (a) two targets, (b) multiple targets and (c) an airplane model.

In conclusion, we have proposed and experimentally demonstrated a chip-based photonic radar based on a silicon photonic platform. Key photonic devices in radar transceiver and receiver are integrated on a single chip, which can generate the frequency–multiplied LFM signal and perform the wideband de-chirp processing required for high resolution imaging radars. A chip-based photonic radar covering the full Ku band is established, which realizes a range resolution of 2.7 cm and a measurement error of less than 2.75 mm. High-resolution ISAR imaging of two and multiple targets and a complex-profile object is successfully implemented. The demonstration reveals that photonics-based radars not only exhibit high performance but can also occupy very small footprint even operated in the low-frequency regime. Since the photonic radar chip is fabricated on a CMOS-compatible platform, it is feasible to integrate both the photonic circuits and the electrical components on the same chip, so the size and weight of the radar can be further reduced.

**REFERENCES**

1. V. C. Chen, and M. Martorella, *Inverse synthetic aperture radar imaging: principles, algorithms and applications* (SciTech Publishing, 2014).
2. G. Franceschetti, R. Lanari, *Synthetic aperture radar processing* (CRC press, 2018).
3. J. Yu, F. Zhao, J. Cali, F. Dai, D. Ma, X. Geng, Y. Jin, Y. Yao, X. Jin, J. Irwin, and R. Jageger, IEEE J. Solid-State Circuit **49**, 1905 (2014).
4. R. Ebelt, A. Hamidian, D. Shmakov, T. Zhang, V. Subramanian, G. Boeck, and M. Vossiek, IEEE Trans. Microw. Theory Tech. **62**, 2193 (2014).
5. G. Pyo, C. Kim and S. Hong, "Single-antenna FMCW radar CMOS transceiver IC," IEEE Trans. Microw. Theory Tech. **65**, 945 (2017).
6. Y. Wang, L. Lou, B. Chen, Y. Zhang, K. Tang, L. Qiu, S. Liu, and Y. Zheng, IEEE Trans. Microw. Theory Tech. **65**, 4385 (2017).
7. Y. Wang, K. Tang, Y. Zhang, L. Lou, B. Chen, S. Liu, L. Qiu, and Y. Zheng, IEEE International Solid-State Circuits Conference (ISSCC) (IEEE, 2016), pp. 240-241.
8. P. Ghelfi, F. Laghezza, F. Scotti, G. Serafino, A. Capria, S. Pinna, D. Onori, C. Porzi, M. Scaffardi, A. Malacarne, V. Vercesi, E. Lazzeri, F. Berizzi and A. Bogoni, Nature, **507**, 341, (2014).
9. R. Li, W. Li, M. Ding, Z. Wen, Y. Li, L. Zhou, S. Yu, T. Xing, B. Gao, Y. Luan, Y. Zhu, P. Guo, Y. Tian, and X. Liang, Opt. Express, **25**, 14334 (2017).
10. F. Zhang, Q. Guo, Z. Wang, P. Zhou, G. Zhang, J. Sun, and S. Pan, Opt. Express, **25**, 16274 (2017).
11. Wang, J. Wo, X. Luo, Y. Wang, W. Cong, P. Du, J. Zhang, B. Zhao, J. Zhang, Y. Zhu, J. Lan, and L. Yu, Opt. Express, **26**, 20708 (2018).
12. S. Zhang, W. Zou, N. Qian, and J. Chen, IEEE Photon. Technol. Lett. **30**, 1028 (2018).
13. S. Peng, S. Li, X. Xue, X. Xiao, D. Wu, X. Zheng, and B. Zhou, Opt. Express **26**, 1978 (2018).
14. X. Ye, F. Zhang, Y. Yang, and S. Pan, Photon. Res. **7**, 265 (2019).
15. Y. Yao, F. Zhang, Y. Zhang, X. Ye, D. Zhu, and S. Pan, Optical Fiber Communication Conference (OFC), (Optical Society of America, 2018), paper Th3G.5.
16. F. Zhang, Q. Guo, Y. Zhang, Y. Yao, Pei. Zhou, D. Zhu, and S. Pan, Chin. Opt. Lett. **15**, 112801 (2017).
17. G. Serafino, F. Scotti, L. Lembo, B. Hussain, C. Porzi, A. Malacarne, S. Maresca, D. Onori, P. Ghelfi, and A. Bogoni, J. Lightw. Technol. **37**, 643 (2019).
18. G. Serafino, F. Amato, S. Maresca, L. Lembo, P. Ghelfi, and A. Bogoni, IET Radar Sonar Navig., **12**, 1179 (2018).
19. S. Melo, S. Maresca, S. Pinna, F. Scotti, M. Khosravanian, A. Jr., F. Giannetti, A. Barman, and A. Bogoni, J. Lightw. Technol. **36**, 2337 (2018).
20. D. Marpaung, J. Yao, and J. Capmany, Nat. Photon., **13**, 80 (2019).
21. J. Fandiño, P. Muñoz, D. Doménech, and J. Capmany, Nat. Photon., **11**, 124 (2016).
22. M. Huang, S. Li, M. Xue, L. Zhao, and S. Pan, Opt. Express, **26**, 23215 (2018).
23. M. Burla, D. Marpaung, L. Zhuang, A. Leinse, M. Hoekman, R. Heideman, and C. Roeloffzen, IEEE Photon. Technol. Lett. **25**, 1145 (2013).
24. W. Zhang, and J. Yao, J. Lightwave Technol. **36**, 4655 (2018).
25. L. Zhuang, C. Roeloffzen, M. Hoekman, K. Boller, and A. Lowery, Optica **2**, 854 (2015).
26. S. L. Pan and J. P. Yao, IEEE Trans. Microw. Theory Tech., **58**, 1967 (2010).